\documentstyle[12pt]{article}
\newcommand{\be}{\begin{equation}}
\newcommand{\ee}{\end{equation}}
\newcommand{\ba}{\begin{eqnarray}}
\newcommand{\ea}{\end{eqnarray}}

\begin{document}
\begin{center}
{\bf EXACTLY SOLVABLE TWO-DIMENSIONAL COMPLEX MODEL WITH REAL
SPECTRUM
}\\
\vspace{0.5cm} {\large \bf F. Cannata$^{1,}$\footnote{E-mail:
cannata@bo.infn.it}, M.V. Ioffe$^{2,3,}$\footnote{E-mail:
m.ioffe@pobox.spbu.ru} ,
D.N. Nishnianidze$^{2,4,}$\footnote{E-mail: qutaisi@hotmail.com}}\\
\vspace{0.2cm}
$^1$ Dipartimento di Fisica and INFN, Via Irnerio 46, 40126 Bologna, Italy.\\
$^2$ Department of Theoretical Physics, Sankt-Petersburg State University,\\
198504 Sankt-Petersburg, Russia\\
$^3$ Departamento de Fisica Teorica, Atomica y Optica, Universidad
de Valladolid,
47071 Valladolid, Spain\\
$^4$ Kutaisi Technical University, 4614 Kutaisi, Republic of
Georgia
\end{center}
\vspace{0.2cm} \hspace*{0.5in}
\hspace*{0.5in}
\begin{minipage}{5.0in}
{\small Supersymmetrical intertwining relations of second order in
derivatives allow to construct a two-dimensional quantum model
with complex potential, for which {\it all} energy levels and
bound state wave functions are obtained analytically. This model
{\it is not amenable} to separation of variables, and it can be
considered as a specific complexified version of generalized
two-dimensional Morse model with additional $\sinh^{-2}$ term. The
energy spectrum of the model is proved to be purely real. To our
knowledge, this is a rather rare example of a nontrivial exactly
solvable model in two dimensions. The symmetry operator is found,
the biorthogonal basis is described, and the pseudo-Hermiticity of
the model is demonstrated. The obtained wave functions are found
to be common eigenfunctions both of the Hamiltonian and of the
symmetry operator. This paper is dedicated
to the eightieth birthday of Yuri Victorovich Novozhilov.\\
\vspace*{0.1cm} PACS numbers: 03.65.-w, 03.65.Fd, 11.30.Pb }
\end{minipage}
\vspace*{0.2cm}
\section*{\bf 1. \quad Introduction}
\vspace*{0.1cm} \hspace*{3ex} While the role of exactly solvable
and partially (quasi-exactly) solvable models in one-dimensional
Quantum Mechanics is well known\footnote{Here and below we call
the system exactly solvable if all eigenvalues and eigenfunctions
of its bound states are known analytically. The system is called
partially (quasi-exactly) solvable when a part of them is known.},
the number of these models is rather restricted. For
two-dimensional systems (exact- and partial-) solvability is a
much harder task. Besides models with separation of variables,
which are actually reduced to couples of one-dimensional models,
only Calogero and the so called Calogero-like
models\cite{calogero} (with the number of particles
$N=3$)\footnote{Incidentally, we would like to mention that like
the Calogero models all considered models (including this
generalized Morse model) cam alternatively be interpreted
\cite{neelov} either as describing three particles on a line or as
a particle in a two-dimensional space.} are exactly solvable. Some
partially solvable systems were constructed in \cite{new},
\cite{pseudo}, \cite{ioffe}, \cite{iv} by the supersymmetrical
method of $SUSY-$separation variables. Just the methods of
supersymmetry seem to be the most appropriate ones to attack this
fundamental problem of constructing exactly and partially solvable
quantum (and classical \cite{classical}) models in the case of few
space dimensions.

The first step in this direction was taken in \cite{david}, where
a list of partial solutions of supersymmetrical intertwining
relations for two-dimensional systems was found. All these models
are integrable, i.e. there is a dynamical symmetry of fourth order
in momenta, such that the corresponding generators are in
involution with the Hamiltonians. After that two new
supersymmetrical methods - of $SUSY$-separation of variables and
of shape invariance -  were proposed \cite{new}, \cite{ioffe},
\cite{iv}, \cite{double} which provided an opportunity to obtain
part of the spectrum and wave functions for some of these
integrable models. Thus a class of partially solvable integrable
two-dimensional quantum systems was built. Naturally, it would be
very important to obtain some {\it exactly} solvable models among
these integrable ones.

In the present paper a new idea, which may be useful in order to
solve the task formulated above, will be presented. Namely, we
will choose a particular value of parameter ($a=-1/2$) in the
two-dimensional Morse potential with additional $\sinh^{-2}$ term
\cite{new} in such a way that one partner Hamiltonian in the SUSY
intertwining relations {\it will allow} for standard separation of
variables. In such a case this Hamiltonian is exactly solvable,
and its spectrum and eigenfunctions will be found analytically.
The second partner Hamiltonian still {\it does not allow for} the
separation of variables, but due to intertwining relations its
spectrum and eigenfunctions can also be obtained analytically by
acting with the supercharge. Since SUSY intertwining relates the
spectra of partner Hamiltonians up to normalizable zero modes of
supercharges, the requirement of exact solvability of both
Hamiltonians amounts to the condition that all zero modes of
supercharges are under control. This could be implemented for the
generalized two-dimensional Morse model (with additional
$\sinh^{-2}$ term), but as will be shown in Section 2, the direct
analysis of singularities of wave functions for $a=-1/2$ case is
rather problematic in a general form due to the presence of
confluent hypergeometric functions. Nevertheless the principal
opportunities offered by the method are demonstrated for a
particular energy level. Consideration in Section 3 of some
specific complex version of this model (which regularizes the
repulsive singularity for the real values of $x_1,x_2$) leads to a
more interesting system: {\it all} bound states and normalizable
wave functions can be derived explicitly. The properties of this
last model are investigated. The most important feature is the
reality of the energy spectrum (which does not depend on the
parameter of complexification $\delta$). In addition, in Section 4
the bound-state-biorthogonal-basis \cite{mostafazadeh} will be
constructed in a natural way, and the action of symmetry operator
will be calculated. In particular, it will be shown that the
obtained wave functions are the common eigenfunctions both of the
Hamiltonian and of the symmetry operators.

\section*{\bf 2.\quad Generalized two-dimensional Morse potential
with $a=-1/2$} \vspace*{0.1cm} \hspace*{3ex} The main element of
the two-dimensional second order Supersymmetrical Quantum
Mechanics (SUSY QM) \cite{david}, \cite{ioffe} is represented in
terms of the intertwining relations: \be \tilde HQ^+ = Q^+H;\quad
Q^-\tilde H = HQ^-;\label{intertw} \ee between a pair of scalar
Hamiltonians $H,\,\tilde H:$ \ba H = -\triangle + V(\vec x);\quad
\tilde H = -\triangle + \tilde V(\vec x); \quad  \triangle \equiv
\partial_1^2 + \partial_2^2; \quad
\partial_i \equiv \partial/\partial x_i; \quad i=1,2.\label{defh}
\ea These intertwining relations realize the isospectrality up to
zero modes of $Q^{\pm}$ of the superpartners $H, \,\tilde H$ and
the connection between their wave functions with the same values
of energy: \be \Psi_n(x) = Q^- \tilde\Psi_n(x);\quad
\tilde\Psi_n(x) = Q^+ \Psi_n(x); \quad n=0,1,2,... \label{psi} \ee

Though the solutions of intertwining relations (\ref{intertw})
have to be searched in principle for the supercharges $Q^{\pm}$
with the most general form of second order differential operators
(see \cite{david}, \cite{ioffe} for details), we will restrict
ourselves to the particular solution of Lorentz (hyperbolic) type:
\ba Q^+ &=& (\partial_1^2 - \partial_2^2) +  C_i \partial_i + B =
4\partial_+\partial_- +C_+\partial_- +
C_-\partial_+ + B; \label{ourq}\\
Q^- &=& (\partial_1^2 - \partial_2^2) -  C_i \partial_i + B =
4\partial_+\partial_- -C_+\partial_- - C_-\partial_+ +
B;\,\,i=1,2. \label{ourqq} \ea Then the potentials $ \tilde V(\vec
x), V(\vec x) $ and the function $ B(\vec x) $ can be expressed in
terms of four functions - $F_1(2x_1),\,F_2(2x_2)$ and
$C_{\pm}(x_{\pm}):$ \ba \tilde V&=&\frac{1}{2}(C_+' + C_-') +
\frac{1}{8}(C_+^2 + C_-^2) +
\frac{1}{4}\biggl( F_2(x_+ -x_-) - F_1(x_+ + x_-)\biggr) ,\nonumber\\
V&=&-\frac{1}{2}(C_+' + C_-') + \frac{1}{8}(C_+^2 + C_-^2) +
\frac{1}{4}\biggl( F_2(x_+ -x_-) - F_1(x_+ + x_-)\biggr) ,\label{potential}\\
B&=&\frac{1}{4}\biggl( C_+ C_- + F_1(x_+ + x_-) + F_2(x_+ -
x_-)\biggr) . \label{functionB} \ea These functions must satisfy
the following equation: $$
\partial_-(C_- F) =
-\partial_+(C_+ F),$$ where $x_{\pm} \equiv x_1\pm x_2\quad
\partial_{\pm}=\partial / \partial x_{\pm} $ and $C_{\pm}$ depend only on
$ x_{\pm},$ respectively:
$$C_+ \equiv C_1 - C_2 \equiv C_+(x_+);\quad
C_- \equiv C_1 + C_2 \equiv C_-(x_-),
$$
and
$$ F=F_{1}(x_{+}+x_{-}) + F_{2}(x_{+}-x_{-}).$$

Among different particular solutions of this class \cite{david},
\cite{ioffe} we will choose here the well studied two-dimensional
generalization of one dimensional Morse potential. The model is
not amenable to standard separation of variables, and it is
defined by: \ba
C_+&=&4a\alpha;\quad C_-=4a\alpha\cdot\coth \frac{\alpha x_-}{2}   \label{cpm}\\
f_1(x_1)&\equiv & \frac{1}{4} F_1(2x_1)=-A\biggl(\exp(-2\alpha
x_1) - 2 \exp(-\alpha x_1)\biggr); \label{f1}\\ f_2(x_2)&\equiv &
\frac{1}{4} F_2(2x_2)= +A\biggl(\exp(-2\alpha x_2) - 2
\exp(-\alpha x_2)\biggr)
\label{f2}\\
\tilde V(\vec x)&=& \alpha^2a(2a-1)\sinh^{-2}\biggl(\frac{\alpha
x_-}{2}\biggr) +
4a^2\alpha^2 +\nonumber\\
 &+&A \biggl[\exp(-2\alpha
x_1)-2 \exp(-\alpha x_1) + \exp(-2\alpha x_2)-2 \exp(-\alpha x_2)\biggr]\label{tildemorse}\\
V(\vec x)&=& \alpha^2a(2a+1)\sinh^{-2}\biggl(\frac{\alpha
x_-}{2}\biggr) +
4a^2\alpha^2 +\nonumber\\
 &+&A \biggl[\exp(-2\alpha
x_1)-2 \exp(-\alpha x_1) + \exp(-2\alpha x_2)-2 \exp(-\alpha
x_2)\biggr], \label{morse} \ea where parameters $a,\,A>0,\,\alpha
>0$  are arbitrary real numbers.

Just on the basis of this model two new methods -
$SUSY$-separation of variables and two-dimensional shape
invariance - were elaborated for the first time \cite{new},
\cite{ioffe}. As a result, partial solvability of the model was
discovered \cite{new}, \cite{ioffe}, \cite{double} and a variety
of wave functions $\Psi_n, \tilde\Psi_n$ was found analitycally
for parameters restricted to the range: \be a \in (-\infty,\,
-\frac{1}{4}-\frac{1}{4\sqrt{2}}); \quad
\frac{\sqrt{A}}{\alpha}-n-\frac{1}{2} > -2a > 0;\quad n=0,1,2,...
, \label{region1} \ee which provides the condition of
normalizability of zero modes of $Q^+ $ and the absence of the
"fall to the centre" (see details in \cite{new},\cite{ioffe}).

It is worth to recall now the main idea of the method of
$SUSY$-separation of variables. By similarity transformation the
supercharge $Q^{\pm}$ (\ref{ourq}) for Lorentz metrics can be
transformed to the operators without linear derivatives:
\ba
q^{\pm} &=& \exp{(-\chi (\vec x))} Q^{\pm} \exp{(+\chi (\vec x))}
= \partial_1^2-\partial_2^2+\frac{1}{4}(F_1(2x_1)+F_2(2x_2));
\label{smallq}\\
\chi (\vec x) &\equiv & -\frac{1}{4}\Bigl(\int C_+(x_+)dx_+ +\int
C_-(x_-)dx_-\Bigr),\nonumber \ea which allow for the separation of
variables (this is why we call this method as $SUSY$-separation).
If $-F_1$ and $+F_2$ belong to a class of exactly solvable
one-dimensional potentials, the normalizable zero modes
$\Omega_n(\vec x)$ of $Q^+$ can be found analytically. Then due to
(\ref{intertw}) a set of wave functions of $H$ can be constructed
as linear combinations of these zero modes, leading therefore to
partial solvability of the model. The shape invariance method
\cite{new} (together with the second shape invariance
\cite{double} of the model) enlarges the part of the spectrum
known analytically.

Now a new method, involving separation of variables in one partner
Hamiltonian (which does not hold for the other partner), will be
formulated. Let us choose the parameters of the same model in such
a way that the Hamiltonian $H$ {\bf does allow} for the standard
procedure of separation of variables. Then the intertwining
relations and knowledge of zero modes of supercharges provides
full information about the partner Hamiltonian $\tilde H,$ which
{\bf does not allow} conventional separation of variables.

Luckily, a suitable choice of the value of the parameter $a=-1/2$
in (\ref{morse}) makes $H$ amenable to separation of variables:
\be H(\vec x) = h_1(x_1)+h_2(x_2)+\alpha^2;\quad h_1(x_1)\equiv
-\partial_1^2-f_1(x_1);\quad h_2(x_2)\equiv
-\partial_2^2+f_2(x_2), \label{hh} \ee and its wave functions with
energies \be E_{n,m}=\epsilon_n+\epsilon_m + \alpha^2, \label{Enm}
\ee where the last term originates from the free term in
(\ref{morse}), can be written as: \be \Psi_{E_{n,m}} =
c_1\eta_n(x_1)\eta_m(x_2)+c_2\eta_m(x_1)\eta_n(x_2);\quad
c_1,c_2=Const \label{psinm} \ee where $\epsilon_{k}$ and
$\eta_k(x)$ solve the exactly solvable one-dimensional problem
with the standard Morse potential in terms of confluent
hypergeometric functions (see \cite{baitman}): \ba
&&\Biggl(-\partial^2 + A \biggl(\exp(-2\alpha x)-2 \exp(-\alpha
x)\biggr)\Biggr)\eta_k(x)=\epsilon_k\eta_k(x)
\label{eta}\\
&&\eta_k = \exp(-\frac{\xi}{2}) (\xi)^{s_k}
\Phi(-k, 2s_k +1; \xi);\quad \xi\equiv \frac{2\sqrt{A}}{\alpha}\exp(-\alpha x); \label{hyper}\\
&&\epsilon_k=-A[1-\frac{\alpha}{\sqrt{A}}(k+1/2)]^2;\quad
s_k=\frac{\sqrt{A}}{\alpha}-k-1/2. \label{epsilon} \ea Thus the
two-dimensional Schr\"odinger problem with Hamiltonian $H$ is
obviously exactly solvable with 2-fold degeneracy for the levels
with $n\neq m.$

Now the next step of our approach is to use the intertwining
relations (\ref{intertw}) in order to obtain, by means of
(\ref{psi}), the wave functions of the Hamiltonian $\tilde H,$
which {\it does not allow} for separation of variables. It is
clear from the explicit form of $Q^+$ that due to the singularity
of $Q^+$ at $x_-\to 0$ the normalizability of functions
$\tilde\Psi$ should be investigated in detail. Though this problem
is not analyzable fully due to the presence of hypergeometric
functions in (\ref{psinm}), particular normalizable eigenfunctions
of $\tilde H$ can be studied. Indeed, taking antisymmetric
$(c_1=-c_2)$ wave functions $\Psi^A_{E_{n,m}}$ of the form
(\ref{psinm}), one can check\footnote{It is necessary to use
elementary relations between confluent hypergeometric functions
(see \cite{baitman}, vol.1, Subsection 6.3.) and some properties
derived in Subsection 4.4. of \cite{new}.} that for $|n-m|=1$ they
are represented as linear combinations of zero modes
$\Omega_k(\vec x)$ of $Q^+$: $$ \Psi^A_{E_{m+1,m}}=-
\Psi^A_{E_{m,m+1}} =\frac{2(s_{m+1}+1)(2s_{m+1}+1)}{2s_{m+1}+m+2}
\Sigma_{k=0}^{k=m}a_{m+1,k}\Omega_k(\vec x ),$$ and therefore have
{\it no partner} bound states at $\tilde H.$ The symmetric
functions $(c_1=c_2)$ from (\ref{psinm}) with $n=m\pm 1$ give
$Q^+\Psi^S_{E_{m\pm 1,m}}$, which are also absent among bound
states of $\tilde H$ due to their nonnormalizable behaviour at
$x_-\to 0.$ Therefore the spectrum of $\tilde H$ definitely {\it
does not} include $E_{m\pm 1,m}.$

The wave function $\Psi^A_{E_{0,2}}$ leads to the (symmetrical)
$\tilde\Psi^S_{E_{0,2}}$, which can be rewritten as: $$
\tilde\Psi^S_{E_{0,2}}\equiv Q^+\Psi^A_{E_{0,2}} =
\alpha^2(2s_2+3)(\xi_1-\xi_2)^2\exp(-\frac{\xi_1+\xi_2}{2})(\xi_1\xi_2)^{s_2}.
$$ This example demonstrates explicitly that a suitable choice of
$c_1,c_2, n, m$ at (\ref{psinm}) can compensate the singularity in
$Q^+$ and provide a normalizable wave function
$\tilde\Psi_{E_{n,m}}$.

In principle, besides the eigenfunctions $\tilde\Psi$ of $\tilde
H$ discussed above, some additional normalizable eigenfunctions
could exist. If so, due to intertwining relations the action of
the operator $Q^-$ onto these functions should either give zeros
or unnormalizable functions $\Psi$. The first option is excluded
by the analysis of zero modes of $Q^-(a=-1/2)\equiv Q^+(a=+1/2),$
taking into account that $a=+1/2$ is outside the range
(\ref{region1}). The second option also does not materialize since
normalizable eigenfunction of $\tilde H$ with potential
(\ref{tildemorse}) for $a=-1/2$ gives $\tilde\Psi\sim x_-^2$ at
$x_-\to 0$. Therefore action of $Q^-$ cannot affect its
normalizability, and we are back to the case of previous
paragraph.

\section*{\bf 3.\quad Complexified model}
\vspace*{0.1cm} \hspace*{3ex} Since the bound state spectrum of
the model $\tilde H$ for $a=-1/2$ turned out to be difficult to
analyze due to singularities of $\tilde V(\vec x)$ and $Q^{\pm}$
at $x_-=0,$ a natural idea is to remove somehow these
singularities. In one-dimensional Quantum Mechanics the  recipe is
well known \cite{buslaev}, \cite{raj}, \cite{ahmed} - to shift the
space coordinate into the complex plane. It means that from this
moment we deal with complex potentials\footnote{An extensive
literature concerning one-dimensional non-Hermitian Hamiltonians
followed the seminal papers \cite{bender} of C.Bender and
S.Boettcher (see for example, \cite{mostafa}).}, in general.

In our two-dimensional situation\footnote{Non-Hermitian models in
two-dimensional Quantum Mechanics were studied in \cite{pseudo}.}
we have to shift $x_-=x_1-x_2,$ therefore one has to violate the
exchange symmetry of the system under $x_1 \leftrightarrow x_2.$
The easiest way is to replace $$ \vec x\to \vec x +
i\vec\delta;\quad \vec\delta=(\delta, 0) $$ (with $\delta$ small
enough, such that $\alpha\delta\in (0, \pi/2)$) removing the
singularities from the real $(x_1,x_2)$ plane. In terms of $\xi$
of Eq.(\ref{hyper}) a phase factor appears: $\xi\to
e^{-i\alpha\delta}\xi .$  As usual, such imaginary shift of $\vec
x$ preserves {\bf the reality of the spectrum} of the
Schr\"odinger operator.

Under this shift not too many changes affect formulas of Section
2. The operator $Q^-(\vec x +i\vec\delta)$ in (\ref{ourqq})
preserves its form, but from now on it is not hermitian conjugate
of $Q^+((\vec x +i\vec\delta)),$ namely
$Q^-=\Bigl((Q^+)^{\dagger}\Bigr)^{\star}= (Q^+)^{t}.$ Functions
(\ref{cpm}) - (\ref{f2}) become complex, and (also complex)
potentials (\ref{tildemorse}), (\ref{morse}) are related now by $$
\tilde V(\vec x+i\vec\delta) = \tilde V^{\star}(\vec
x-i\vec\delta)=\exp{(-2i\vec\delta\vec\partial)} \tilde
V^{\star}(\vec x+i\vec\delta) \exp{(+2i\vec\delta\vec\partial)}.
$$ This equality expresses the property of pseudo-Hermiticity for
non-Hermitian Hamiltonians, which guarantees in general
\cite{mostafazadeh} that the spectrum consists of real eigenvalues
and complex conjugated pairs. In particular, in our case the whole
bound states spectrum of $\tilde H$ is known to be real.

The eigenfunctions $\eta_k(x)$ of one-dimensional Morse equation
(\ref{psinm}) are expressed in terms of confluent hypergeometric
functions as in (\ref{hyper}), but with $\xi\to
e^{-i\alpha\delta}\xi ,$ and are still normalizable. One can check
that {\it no additional normalizable} solution of (\ref{psinm})
appears in the complex $x$- plane. Indeed, after the substitution
$\eta_k\equiv \exp(-\xi /2)\xi^s Y(\xi),$ the Eq.(\ref{eta}) in
the variable $\xi$ is reduced to the confluent hypergeometrical
equation \cite{baitman}. There are different ways to represent the
general solution of this equation. A convenient one is: $$ Y(\xi)
= c_1 y_5(\xi) + c_2 y_7(\xi),$$ where $\xi$ includes a phase
factor, and definitions and terminology for linearly independent
solutions $y_5, y_7$ are given in \cite{baitman} (see Vol. 1,
Subsection 6.7.): \be y_5=\Phi(a,b;\xi); \quad y_7=\exp
(\xi)\Psi(b-a,b;-\xi). \label{y5y7} \ee Just the exponential in
the second term in (\ref{y5y7}) allows to prove that the only kind
of {\it normalizable} solutions $Y,$ even for our complex $\xi ,$
corresponds to $c_2=0$. Therefore, the conditions of
normalizability lead to Eqs.(\ref{hyper}), (\ref{epsilon}), and
the whole bound state spectrum (\ref{Enm}) of Hamiltonian $H(\vec
x +i\vec\delta)$ is known and it remains {\it real} after this
complexification.

Using again the intertwining relations (\ref{intertw}), one can
obtain the eigenfunctions $$ \tilde\Psi_{E_{n,m}}(\vec x
+i\vec\delta)=Q^+\eta_n(x_1+i\delta)\eta_m(x_2).$$ Due to absence
of singularity of $Q^+$ at $x_-\to 0$, these wave functions are
normalizable.

One has to remember that the partner Hamiltonians $\tilde H$ and
$H$ are isospectral only up to zero modes of supercharges.
Normalizable zero modes of $Q^+$ were constructed and investigated
in detail in \cite{new}. It was proved that a variety of linear
combinations of these zero modes can be built, which are
eigenfunctions of the Hamiltonian $H.$ In particular, this can be
done for the case $a=-1/2$ (when $H$ allows separation of
variables), and the corresponding eigenvalues are \cite{new}: \be
E_k=2\epsilon_k + 2\alpha^2s_k=-2\alpha^2s_k(s_k-1)=
\epsilon_{k+1}+\epsilon_k+\alpha^2;\quad
s_k=\frac{\sqrt{A}}{\alpha}-k-1/2 . \label{Ek} \ee For arbitrary
$n,m$ eigenvalues (\ref{Enm}) of $H$ are two-fold degenerate:
there are symmetrical and antisymmetrical components in
(\ref{psinm}). But for the particular case of $n=m\pm 1$ {\it the
antisymmetrical} combination $\Psi^A_{E_{m\pm 1,m}}$ is
annihilated by $Q^+$. In a contrast to the Hermitian model of
Section 2, for {\it the symmetrical} component $\Psi^S_{E_{m\pm
1,m}}$ its partner state $\tilde\Psi^A_{E_{m\pm
1,m}}=Q^+\Psi^S_{E_{m\pm 1,m}}$ for the complex model has no
singularity at $x_-\to 0,$ and therefore these energy levels
$E_{m\pm 1,m}$ exist in the spectrum of $\tilde H,$ but they are
not degenerate.

The possible zero modes of $Q^-(a)=Q^+(-a)$ can be also
investigated by the same method as in \cite{new}, the
corresponding eigenvalues are $E_k=-2\alpha^2s_k+2\epsilon_k.$
Because these eigenvalues coincide with $E_{k,k-1}$ from the
general expression (\ref{Enm}), no new eigenstates can appear for
$\tilde H$ due to zero modes of $Q^-(a).$


Thus the complex model with the Hamiltonian $\tilde H(\vec
x+i\delta)$ for $a=-1/2$ is {\bf exactly solvable}, its spectrum
is real: \be E_{n,m}=\epsilon_n+\epsilon_m+\alpha^2. \label{final}
\ee For $n=m\pm 1$ these levels are not degenerate, but for all
other $n,m$ there is 2-fold degeneracy. The wave functions of
$\tilde H$ are: \be \tilde\Psi_{E_{n,m}}=Q^+\Psi_{E_{n,m}},
\label{eigen} \ee where $\Psi_{E_{n,m}}$ are given by
(\ref{psinm}) with $\vec x\to \vec x + i\vec\delta;\quad
\vec\delta=(\delta, 0).$

\section*{\bf 4.\quad Integrability and biorthogonality}
\vspace*{0.1cm} \hspace*{3ex} It is known \cite{david},
\cite{ioffe} that "by construction" all Hamiltonians $\tilde H,
H,$ which are intertwined according to (\ref{intertw}) by second
order operators $Q^{\pm},$ are in involution with operators of
fourth order in derivatives: \be \tilde R=Q^+Q^-;\quad
R=Q^-Q^+;\quad [H, R]=0;\quad [\tilde H, \tilde R]=0.
\label{symmetry} \ee These symmetry operators\footnote{Apart from
the case of Laplacian metrics in supercharges, which is not
considered here (see details in \cite{david}).} are not reducible
to functions of Hamiltonians, and therefore all these systems,
including the one presented in this paper, are {\bf integrable}.

In the particular case $a=-1/2$ with separation of variables in
$H,$ considered in this paper, the expressions for $R$ and $\tilde
R$ can be transformed by using the similarity transformation
(\ref{smallq}) and the specific form of one-dimensional
Hamiltonians $h_1(x_1), h_2(x_2)$ in (\ref{hh}): $$
Q^{\pm}=\exp(\pm\chi)q^{\pm}\exp(\mp\chi)=\exp(\pm\chi)(h_2-h_1)\exp(\mp\chi).$$
Then the symmetry operator $R$ for the Hamiltonian with separation
reads: $$ R=(h_2-h_1)^2+2\alpha^2(h_1+h_2)+\alpha^4,$$ and its
wave functions (\ref{psinm}) are simultaneously eigenfunctions of
$R$ with eigenvalues:
$$r_{n,m}=(\epsilon_m-\epsilon_n)^2+2\alpha^2(\epsilon_m+\epsilon_n)+\alpha^4 .$$

The wave functions (\ref{eigen}) of the Hamiltonian $\tilde H$
without separation of variables are also common eigenfunctions
both of $\tilde H$ and $\tilde R:$ $$ \tilde
R\tilde\Psi_{E_{n,m}}=Q^+Q^-Q^+\Psi_{E_{n,m}}=r_{n,m}\tilde\Psi_{E_{n,m}}.$$
Thus the property, noticed in \cite{new} for a few known wave
functions, is fulfilled now for all wave functions in the case of
the exactly solvable model, that we have constructed in this
paper. Though almost all (for $n\neq m$) wave functions
$\tilde\Psi_{E_{n,m}}$ are 2-fold degenerate, the symmetry
operator $\tilde R$ does not mix these degenerate functions.

The factorized wave functions $\Psi_{E_{n,m}}(\vec x+i\vec\delta)$
of $H(\vec x+i\vec\delta)$ and their complex conjugate functions
$\Psi^{\star}_{E_{n,m}}$ form the so called biorthogonal basis for
the non-Hermitian Hamiltonian. The corresponding biorthogonality
relations \ba
&&<\Psi^{\star}_{E_{n,m}}\mid \Psi_{E_{n',m'}}> =
\int d^2x \Psi_{E_{n,m}}(\vec x+i\vec\delta)\cdot
\Psi_{E_{n',m'}}(\vec x+i\vec\delta)
=\nonumber\\
&&=\int dx_1 \eta_n(x_1+i\delta)\cdot \eta_{n'}(x_1+i\delta) \int
dx_2 \eta_n(x_2)\cdot \eta_{n'}(x_2)=\delta_{nn'}\delta_{mm'}
\label{ortho} \ea can either be checked straightforwardly or by
comparing this integral along the line $x_1+i\delta$ with the
analogous integral along the real $x_1$ for the case of $\delta
=0$ and real-valued wave functions $\eta_n$. Absence of
singularities in the narrow strip between these lines means that
relations (\ref{ortho}) follow from the standard orthogonality of
wave functions for the Hermitian Hamiltonians. On the contrary, if
one would consider the scalar product defined with the complex
conjugation of the first multiplier in the integral in
(\ref{ortho}), the connection with the Hermitian case would be not
so evident.

The bound-state-biorthogonal-basis for the non-Hermitian operator
$\tilde H(\vec x+i\vec\delta)$ consists of $\tilde
\Psi_{E_{n,m}}=Q^+ \Psi_{E_{n,m}}$ and
$\tilde\Psi^{\star}_{E_{n,m}}=(Q^+)^{\star}\Psi^{\star}_{E_{n,m}}$.
Due to the equality $Q^-=((Q^+)^{\dagger)^{\star}}$ the scalar
products can be written as: $$ <\tilde\Psi^{\star}_{E_{n,m}}\mid
\tilde\Psi_{E_{n',m'}}> = <(Q^+)^{\star}\Psi^{\star}_{E_{n,m}}\mid
Q^+\Psi_{E_{n',m'}}>= <\Psi^{\star}_{E_{n,m}}\mid Q^-Q^+
\Psi_{E_{n',m'}}>. $$ Since $\Psi_{E_{n',m'}}$ is an eigenfunction
of the symmetry operator $R=Q^-Q^+$ with an eigenvalue $r_{n.m},$
biorthogonality for $\tilde H(\vec x+i\vec\delta)$ follows
directly from (\ref{ortho}), thus leading to a diagonalization of
$\tilde H$ in the bound state subspace.

\section*{\bf Acknowledgements}
The work was partially supported by INFN, the University of
Bologna (M.V.I. and D.N.N.) and by the Spanish MEC (M.V.I. - grant
SAB2004-0143). M.V.I. is grateful to J.Negro and L.M.Nieto for
kind hospitality at Valladolid and useful discussions.

\end{document}